\documentclass[aip,reprint,graphicx]{revtex4-1}
\usepackage{geometry}
\usepackage{setspace}
\usepackage{fullpage}
\usepackage{color}
\usepackage{hyperref}
\usepackage[pdftex]{graphicx}
\newcommand{\beginsupplement}{%
        \setcounter{table}{0}
        \renewcommand{\thetable}{S\arabic{table}}%
        \setcounter{figure}{0}
        \renewcommand{\thefigure}{S\arabic{figure}}%
     }
\draft 

\begin{document}


\title{Anisotropy of the transport properties of NdFeAs(O,F) thin films grown on vicinal substrates} 

\author{Kazumasa\,Iida}\email{iida@mp.pse.nagoya-u.ac.jp}
\affiliation{Department of Materials Physics, Nagoya University, Furo-cho, Chikusa-ku, Nagoya 464-8603, Japan}
\affiliation{JST CREST, Sanbancho 5, Chiyoda-ku, Tokyo 102-0075, Japan}
\author{Takuya\,Matsumoto}
\affiliation{Department of Materials Physics, Nagoya University, Furo-cho, Chikusa-ku, Nagoya 464-8603, Japan}
\author{Keisuke\,Kondo}
\affiliation{Department of Materials Physics, Nagoya University, Furo-cho, Chikusa-ku, Nagoya 464-8603, Japan}
\author{Takafumi\,Hatano}
\affiliation{Department of Materials Physics, Nagoya University, Furo-cho, Chikusa-ku, Nagoya 464-8603, Japan}
\author{Hiroshi\,Ikuta}
\affiliation{Department of Materials Physics, Nagoya University, Furo-cho, Chikusa-ku, Nagoya 464-8603, Japan}


\date{\today}

\begin{abstract}
NdFeAs(O,F) thin films having different fluorine contents were grown on 5$^\circ$ or 10$^\circ$ vicinal cut MgO and CaF$_2$ single crystalline
substrates by molecular beam epitaxy. Structural characterisations by reflection high-energy electron diffraction and x-ray diffraction confirmed the 
epitaxial growth of NdFeAs(O,F). The resistivities of the $ab$-plane and along the $c$-axis ($\rho_{ab}$ and $\rho_{c}$) were derived from the resistivity measurements in the longitudinal and transversal directions. The $c$-axis resistivity was always higher than $\rho_{ab}$, resulting from the anisotropic electronic structure. The resistivity anisotropy $\gamma_ \rho=\rho_c/\rho_{ab}$ at 300\,K was almost constant in the range of $50\leq \gamma_\rho \leq 90$ irrespective of the fluorine content. On the other hand, $\gamma_ \rho$ at 56\,K showed a strong fluorine dependence: $\gamma_ \rho$ was over 200 for the films with optimum fluorine contents (superconducting transition temperature $T_{\rm c}$ around 50\,K), whereas $\gamma_\rho$ was around 70 for the films in the under-doped regime ($T_{\rm c}$ between 35 and 45\,K). The mass anisotropy $\gamma_m=\sqrt{m^*_c/m^*_{ab}}$ ($m^*_c$ and $m^*_{ab}$ are the effective masses along the $c$-axis and on the $ab$-plane) close to $T_{\rm c}$ derived from the anisotropic Ginzburg-Landau approach using the angular-dependency of $\rho_{ab}$ was in the range from 2 to 5. On the assumption $\gamma_m^2=\gamma_ \rho$, those values are small compared to the normal state anisotropy.
\end{abstract}

\maketitle
\section{Introduction}
After the discovery of Fe-based superconductors (FBS)\,\cite{Kamihara}, many efforts have been devoted to searching for new FBS, resulting in the discovery of various new superconductors such as FeSe\cite{Hsu}, (Ba,K)Fe$_2$As$_2$\cite{Rotter} and LiFeAs\cite{Wang}.
To date SmFeAs(O,F) shows the highest $T_{\rm c}$ of 58\,K\,\cite{Fujioka} except for mono-layer FeSe\,\cite{Ge}. The common structural feature of FBS is the FeAs and Fe$Ch$ ($Ch$: chalcogen) tetrahedron, which is believed to play an important role for the superconductivity.
$Ln$FeAs(O,F) ($Ln$: lanthanoide), $Ae$Fe$_2$As$_2$ ($Ae$: alkali earth elements) and  $A$FeAs ($A$: Li and Na) are formed with alternating layers of $Ln$O/FeAs, $Ae$/FeAs and $A$/FeAs, and Fe$Ch$ is composed of a stack of Fe$Ch$ tetrahedra layer. Hence, FBS are expected to be anisotropic in their electronic properties, which may be closely connected with the superconductivity of these compounds. Therefore, investigating the electronic anisotropy may gain an important clue for the mechanism of high-$T_{\rm c}$ superconductivity.

To date, large single crystals of some of the FBS systems have been grown and the temperature dependence of in-plane and out-of-plane resistivity [$\rho_{ab}(T)$ and $\rho_{c}(T)$] and the resultant resistivity anisotropy $\gamma_\rho(T)$ have been measured. A single crystal of FeSe$_{1-x}$Te$_x$ ($x$=0.6) with $T_{\rm c}$=14.2\,K grown by a Bridgeman method showed a metallic behaviour for $\rho_{ab}$ below 140\,K\,\cite{Noji}, whilst it was almost constant above 140\,K. On the other hand, $\rho_{c}$ increased gradually with decreasing temperature down to around 100\,K and then decreased almost linearly with decreasing $T$. The resistivity anisotropy $\gamma_ \rho$ was 44 and 70 at 290\,K and $T_{\rm c}$, respectively.

Song $et$ $al$. reported on the out-of-plane and in-plane resistivity for a LiFeAs single crystal with $T_{\rm c}$=19.7\,K grown by a Bridgeman method\,\cite{Song}. Below 200\,K, $\rho_{ab}(T)$ and $\rho_{c}(T)$ decreased with lowering $T$, whilst the slopes of both resistivity curves decreased close to $T_{\rm c}$. $\gamma_ \rho$ increased from 1.3 at 300\,K to 3.3 at $T_{\rm c}$.

For the BaFe$_2$As$_2$ systems, extensive electrical resistivity measurements on electron, hole and isovalent 
doped single crystals with various doping levels have been reported\,\cite{Tantar-1, Tantar-2, Tantar-3}. The $ab$-plane resistivity for all optimally-doped BaFe$_2$As$_2$ single crystals decreased linearly from 300\,K down to $T_{\rm c}$. Similarly, the resistivity along the $c$-axis also showed a metallic behaviour at low temperature. However, the temperature range of the $T$-linear dependence was different among the BaFe$_2$As$_2$ systems: for P-doped BaFe$_2$As$_2$, $\rho_{c}(T)$ was close to linear in the whole temperature range (i.e. from 300\,K to down to $T_{\rm c}$). On the other hand, $\rho_{c}(T)$ for Co- and K-doped BaFe$_2$As$_2$ showed a broad maximum around 100\,K and 220\,K, respectively.

On the other hand, the size of available $Ln$FeAsO single crystals is still limited. Only three papers  - to the best of our knowledge - regarding the temperature-dependence of $\gamma_ \rho$, PrFeAsO$_{0.7}$\,\cite{Kashiwaya}, SmFeAs(O,F)\,\cite{Moll} and SmFeAsO$_{0.9}$H$_{0.1}$\,\cite{Iimura}, have been published to date due to the difficulty in the crystal growth of this system. Additionally, no studies of the doping dependence of $\gamma_ \rho$ have been reported.
Because the dimensions of the obtained single crystals were typically $\sim200\times200\times10$\,$\mu{\rm m}^3$, a tiny bar was formed for resistivity measurements in both crystallographic main directions by a focused ion beam technique. For all three compounds, the $ab$-plane resistivity decreased almost linearly down to $T_{\rm c}$, whereas along the $c$-axis the resistivity increased continuously to $T_{\rm c}$, which is different from the results reported for other FBS. The respective resistivity anisotropy $\gamma_ \rho$ at 50\,K for PrFeAsO$_{0.7}$, SmFeAs(O,F) and SmFeAsO$_{0.9}$H$_{0.1}$ were 120, 8.4 and 7.8.

As stated above, the anisotropic behaviour of the transport properties of FBS has been investigated using single crystals. 
However, the number of studies on $Ln$FeAsO systems are limited and the obtained results are different from each other. Additionally, no studies of the doping dependence of $\gamma_ \rho$ have been reported.
Here, we report on the transport properties of NdFeAs(O,F) thin films with different F-doping levels grown on vicinal substrates for which the [001] direction is 5$^\circ$ or 10$^\circ$ tilted toward the [100] direction. Using these off-axis grown thin films, we evaluate the resistivity anisotropy, adopting the method that has been employed to study the anisotropy of Bi$_2$Sr$_2$CaCu$_2$O$_8$\,\cite{Zahner}, YBa$_2$Cu$_3$O$_7$ (YBCO)\,\cite{Haage, Czerwinka, Emergo}, MgB$_2$\,\cite{Polyanskii} and Fe(Se,Te)\,\cite{Bryja}. 

\section{Experiment}
\begin{figure}[ht]
	\centering
		\includegraphics[width=\columnwidth]{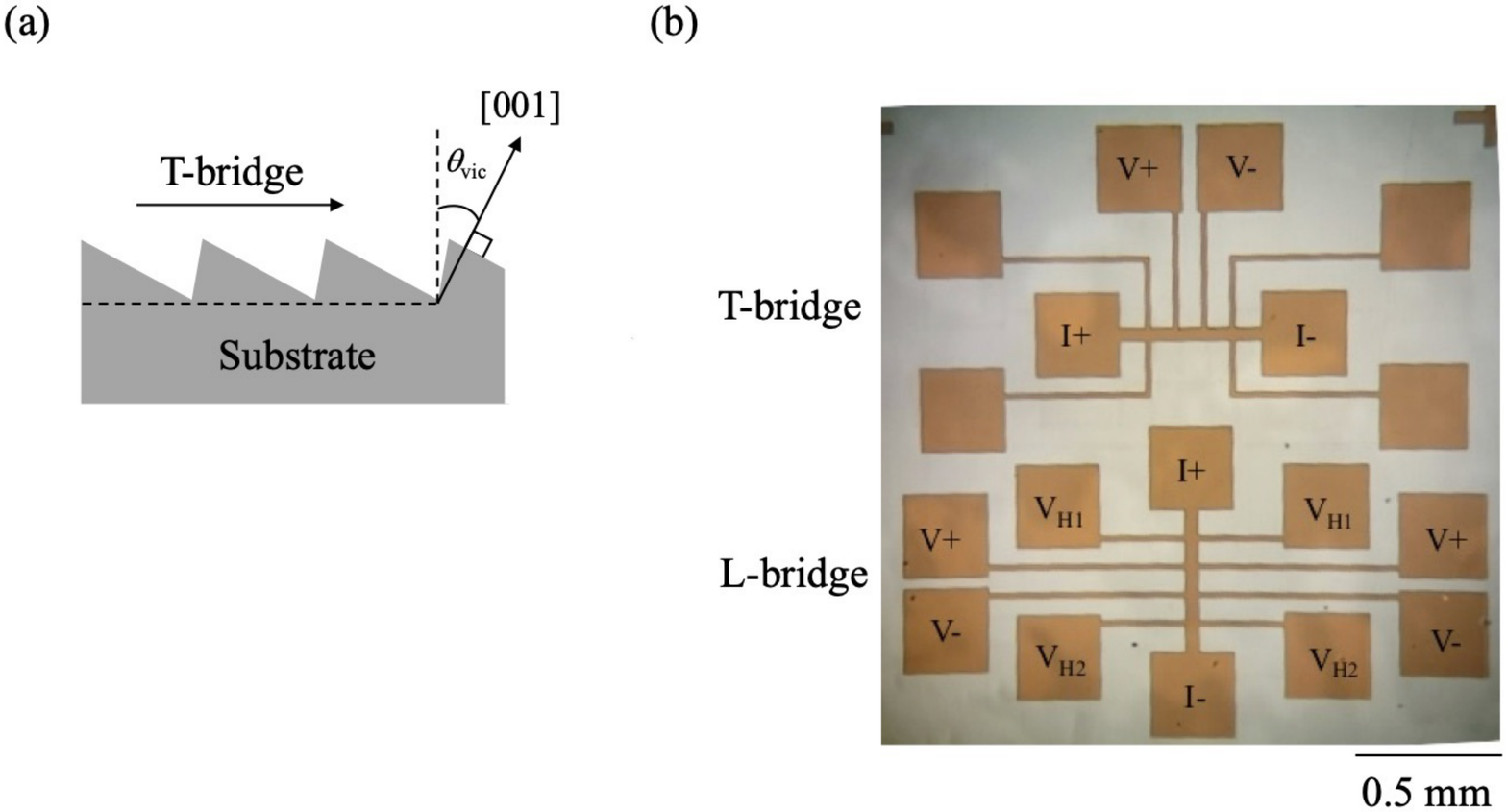}
		\caption{(a) Schematic illustration of a vicinal substrate for which the [001] direction is away from the substrate normal by $\theta_{\rm vic}$. (b) An example of a top-view optical micrograph for the micro-bridges. A bridge running transverse to the vicinal steps is defined as a T-bridge, whereas a bridge parallel to the steps is defined as a L-bridge.} 
\label{fig:figure1}
\end{figure}

Parent NdFeAsO thin films having a thickness of 28--90\,nm were grown on vicinal cut MgO(001) and CaF$_2$(001) single crystalline substrates at 800$^\circ$C by molecular beam epitaxy (MBE). The nominal vicinal angle $\theta_{\rm vic}$ was 5$^\circ$ or 10$^\circ$ measured from the substrate normal toward the [001] directions, as shown in fig.\,\ref{fig:figure1}(a). The film growth was monitored by reflection high-energy electron diffraction (RHEED). To obtain NdFeAs(O,F) films with different F content, a 20\,nm-thick NdOF over-layer was deposited on top of the NdFeAsO layer at various temperatures in the range of $550^\circ{\rm C}\leq T_{\rm dep} \leq 800^\circ$C\,\cite{Kawaguchi-1, Iida-1}, where $T_{\rm dep}$ is the deposition temperature.  Since it is difficult to determine the fluorine content precisely, both the $c$-axis lattice parameter and $T_{\rm c}$ have been used as the indicators of the F-content in NdFeAs(O,F). Additionally, the carrier concentration of some of the films was measured by Hall effect. To rule out the possibility that the NdOF over-layer affected the transport properties, NdOF was removed for some of the NdFeAs(O,F) films by Ar-ion beam etching. We confirmed that the presence or absence of the NdOF over-layer gives no difference in the transport properties of NdFeAs(O,F) (Supplementary figure\,\ref{fig:figureS1}).

Phase purity and the out-of-plane texture were measured by x-ray diffraction (XRD) in Bragg-Brentano geometry using Cu-K$\alpha$ radiation. The growth angle $\alpha$ (i.e. offset angle) was determined as the tilt angle where the intensity maxima was observed for the 003 reflection. In-plane orientation of NdFeAs(O,F) was investigated by $\phi$-scans of the 200 peak.

\begin{table*}[bt]
\centering
\caption{\label{tab:table1}Sample name, $T_{\rm dep}$ of NdOF, the ratio of the thickness of NdOF ($d_{\rm NdOF}$) and NdFeAsO ($d_{\rm NdFeAsO}$), the $c$-axis lattice parameter, the growth angle (offset angle) and the onset $T_{\rm c}$ ($T_{\rm c}^{\rm onset}$) of the samples used in this study. The sample nomenclature is based on the combination of the doping level, the value of onset $T_{\rm c}$ and the substrate. ``OP" represents the optimum doping level at which the NdFeAs(O,F) films show $T_{\rm c}$ higher than 46\,K. The NdFeAs(O,F) films having $T_{\rm c}$ below 44\,K are defined as ``UD". ``PC" represents the parent compound, NdFeAsO.}

\begin{ruledtabular}
\begin{tabular}{cccccc}
Sample name & $T_{\rm dep}$ ($^\circ$C) & $d_{\rm NdOF}$/$d_{\rm NdFeAsO}$ & $c$-axis\,(nm) & Growth angle $\alpha(^\circ)$ & $T_{\rm c}^{\rm onset}$(K) \\ \hline
PC/MgO          &-     &0/45    &0.8590   &12.25&-\\
PC/CaF$_2$   &-     &0/28    &0.8669   &5.17  &-\\
UD0/MgO         &550&20/30  &0.8591   &6.28   &-\\
UD0/CaF$_2$  &700&20/90  &0.8681   &5.23   &-\\
UD36/MgO       &650&20/90  &0.8576   &6.91&36\\
UD42/MgO       &650&20/30  &0.8586   &5.80&42\\
UD43/MgO       &750&20/90  &0.8569   &6.85&43\\
UD44/CaF$_2$&800&20/90  &0.8663   &5.18&44\\
OP45/MgO       &800&20/30  &0.8533   &5.23&45\\
OP46/MgO       &800&20/30  &0.8535   &11.68&46\\
OP49/MgO       &800&20/90  &0.8540    &6.33&49\\
OP56/CaF$_2$&800&20/90  &0.8619    &5.06&56\\
\end{tabular}
\end{ruledtabular}
\end{table*}

After structural characterisations, the films were photolithographically patterned and etched by Ar-ion milling to form micro-bridges for transport measurements. 
As shown in fig.\,\ref{fig:figure1}(b), the bridges were designed to measure the resistivity in the longitudinal direction (abbreviated as L-bridge and the corresponding resistivity $\rho_{\rm L}$) and in the transversal direction (abbreviated as T-bridge and the corresponding resistivity $\rho_{\rm T}$). In the longitudinal direction, the bias current flows within the $ab$-plane only (i.e. $\rho_{\rm L}$=$\rho_{ab}$), whereas in the transversal direction it flows within the $ab$-plane as well as along the $c$-axis.  
Therefore, the $c$-axis resistivity can be calculated by the following equation\,\cite{Zahner},

\begin{equation}
\rho_{c}=(\rho_{\rm T}-\rho_{\rm L}{\rm cos}^2\alpha)/{\rm sin}^2\alpha.
\end{equation}

\noindent
A precision of 0.005$^\circ$/2$\theta$ is guaranteed for our XRD device, which yields a relative uncertainty of $\left| \frac{\delta \rho_c}{\rho_c} \right|$$\sim$0.05\% estimated using the typical values of $\rho_{\rm T}$ and $\rho_{\rm L}$ of our films. 
Misalignment may also occur when the sample was mounted on the sample holder. We checked the XRD data of our recent films grown on ordinary single crystalline substrates and found that the misalignment angle was at most 0.020$^\circ$/2$\theta$. This yields a relative uncertainty of $\left| \frac{\delta \rho_c}{\rho_c} \right|$$\sim$0.39\%. Together, the maximum uncertainty due to the inaccuracy of $\alpha$ is $\left| \frac{\delta \rho_c}{\rho_c} \right|$$\sim$0.50\% in our measurements.
 The bridges had dimensions 50\,$\mu{\rm m}$-wide and 0.5\,mm-long. 
For measuring the angular-dependence of $\rho_{ab}$, a magnetic field, $H$, was applied in the maximum Lorentz force configuration ($H$ perpendicular to the bias current) at an angle $\theta$ measured from the $c$-axis. For some of the NdFeAs(O,F) films, Hall resistance was measured in the field range of $\mu_0H=\pm9$\,T to evaluate the carrier concentration. The samples studied here are summarised in table\,\ref{tab:table1}.

\section{Results and discussion} 
\begin{figure}[ht]
	\centering
		\includegraphics[width=\columnwidth]{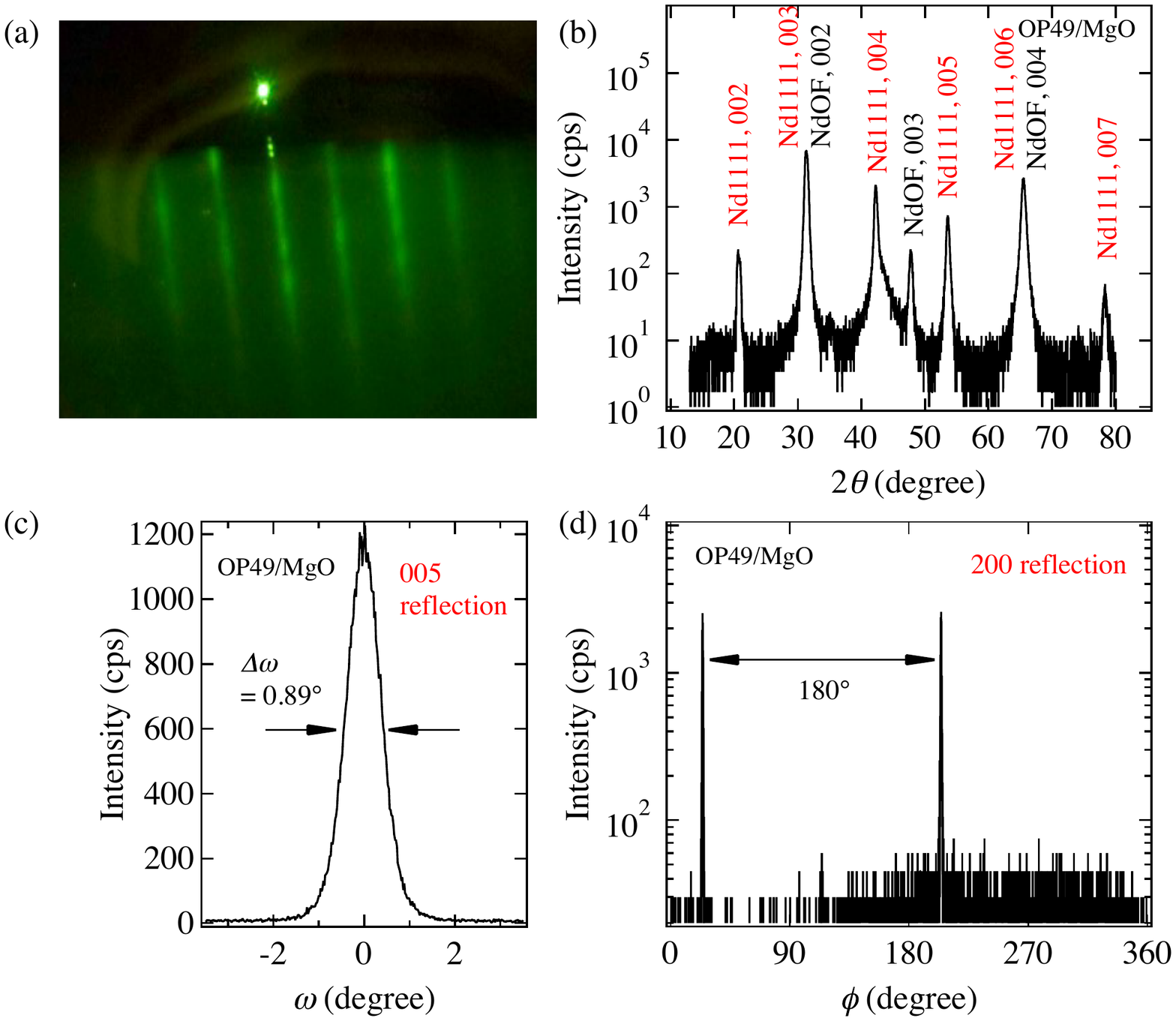}
		\caption{(a) A representative RHEED image of NdFeAsO during the film growth. (b) The $\theta/2\theta$-scan (Cu-K$\alpha$) of the NdFeAs(O,F) (Nd1111) thin film grown on a vicinal MgO substrate measured with an offset angle of 6.33$^\circ$. (c) The 005 rocking curve and (d) the 200 $\phi$-scan of the film shown in fig.\,(b)} 
\label{fig:figure2}
\end{figure}

The RHEED image of NdFeAsO during the film growth of OP49/MgO showed long streaky patterns tilted by $\sim$5$^\circ$, indicative of an epitaxial growth with a smooth surface [fig.\,\ref{fig:figure2}(a)]. This RHEED pattern maintained until the termination of the growth of NdFeAsO. We confirmed that all RHEED images of the NdFeAsO layers showed streaky patterns irrespective of the kinds of substrates as well as the vicinal angles. Figure\,\ref{fig:figure2}(b) shows the XRD pattern of the corresponding film presented in fig.\,\ref{fig:figure2}(a) measured with an offset angle of 6.33$^\circ$, at which the diffraction intensity of the 003 reflection was maximum (the nominal vicinal angle=5$^\circ$). The deposition temperature of NdOF was 800$^\circ$C. Due to the offset angle, the data below 13$^\circ$ were absent. Pronounced 00$l$ diffraction peaks from NdFeAs(O,F) and NdOF were observed, indicating $c$-axis orientation for both compounds. The $c$-axis lattice parameter of NdFeAs(O,F) calculated using the Nelson-Riley function\,\cite{Nelson} was 0.8540\,nm, which is smaller than the parent NdFeAsO film on MgO substrate ($c$=0.8590\,nm, see table\,\ref{tab:table1}). The shortening of the $c$-axis length indicates that F was doped into NdFeAsO, resulting in the formation of the NdFeAs(O,F) phase.
Indeed, the Hall measurements revealed an increase in the carrier density $n$ for OP49/MgO compared from that of parent NdFeAsO: $n$ at 50\,K for OP49/MgO was 1.98$\times$10$^{21}$\,cm$^{-3}$ when estimated using a single carrier model, whereas the corresponding value for NdFeAsO was 0.02$\times$10$^{21}$\,cm$^{-3}$, although the precise evaluation of $n$ is difficult due to the multi-band nature of FBS. The rocking curve of the 005 reflection for OP49/MgO showed a full width at half maximum $\Delta \omega$ of 0.89$^\circ$, indicative of a good out-of-plane texture [fig.\,\ref{fig:figure2}(c)]. The 200 $\phi$-scan of the NdFeAs(O,F) film exhibited two peaks separated by 180$^\circ$ [fig.\,\ref{fig:figure2}(d)], which differs from an epitaxial NdFeAs(O,F) film on an ordinary MgO substrate. This is because the rotational axis for $\phi$-scan is tilted away from the crystallographic $c$-axis. These results proved that the film was epitaxially grown with the $c$-axis tilted by 6.33$^\circ$. We confirmed that all films investigated in this study were epitaxially grown.

\begin{figure}[ht]
	\centering
		\includegraphics[width=\columnwidth]{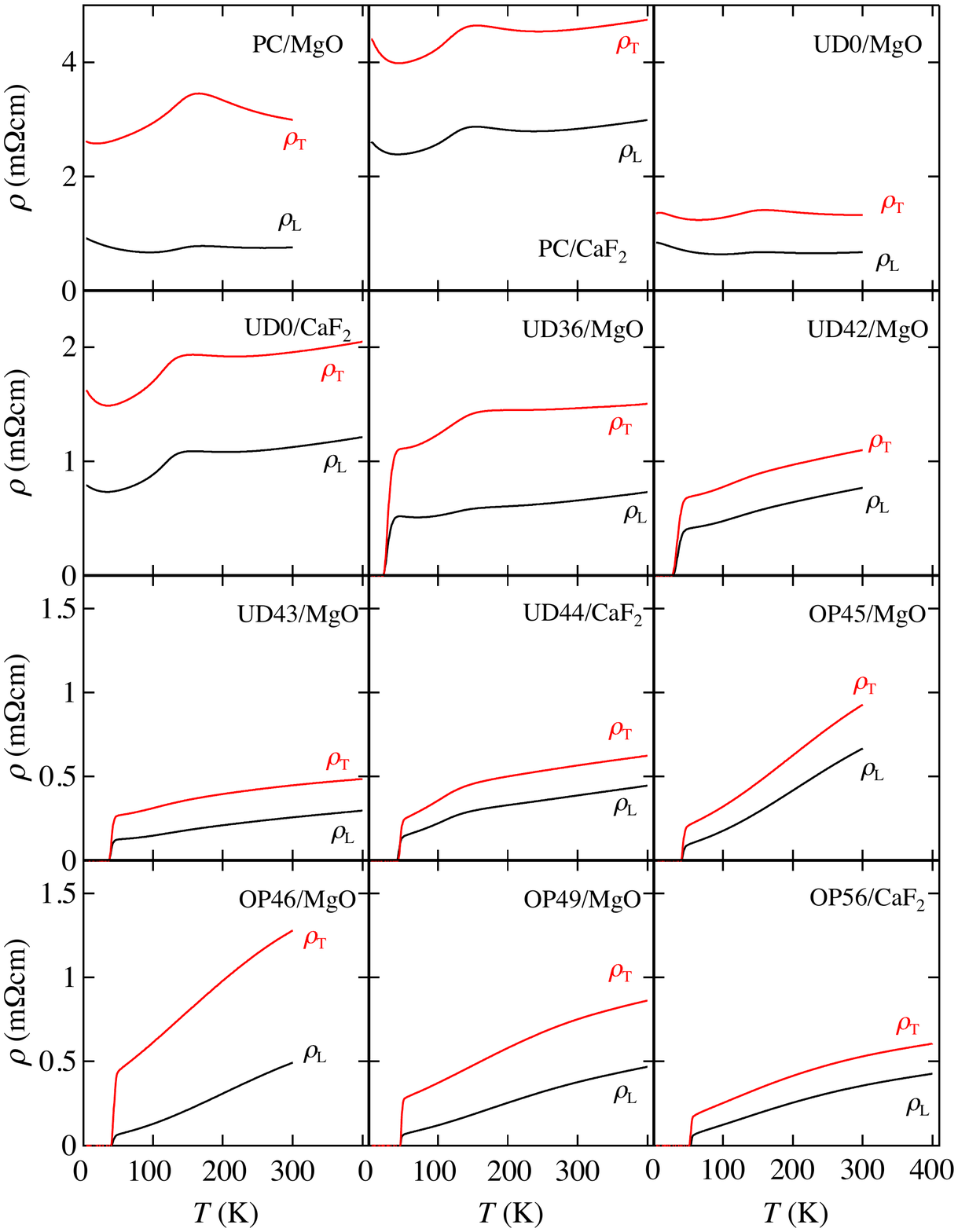}
		\caption{The resistivity curves for all samples tabulated in table\,\ref{tab:table1} along the transversal and longitudinal directions [$\rho_{\rm T}(T)$ and $\rho_{\rm L}(T)$].} 
\label{fig:figure3}
\end{figure}

Figure\,\ref{fig:figure3} summarises the temperature dependence of the resistivity in the transversal and longitudinal directions [$\rho_{\rm T}(T)$ and $\rho_{\rm L}(T)$] for all samples shown in table\,\ref{tab:table1}.
The measured resistivity in the transversal direction $\rho_{\rm T}(T)$ was always higher than in the longitudinal direction $\rho_{\rm L}(T)$ due to the anisotropic electronic structure of NdFeAs(O,F). For the superconducting thin films, both transversal and longitudinal bridges had almost the same $T_{\rm c}$, proving that all films were homogeneous. Another distinct feature is that $\rho_{\rm T}(T)$ increased with increasing the vicinal angle due to the increase of the $c$-axis component. This can be clearly seen in the normalised resistivity traces for OP45/MgO and OP46/MgO (Supplementary figure\,\ref{fig:figureS2}). As can be seen, the difference between $\rho_{\rm T}$ and $\rho_{\rm L}$ for OP45/MgO is larger than that for OP46/MgO because of the larger tilt angle.

\begin{figure}[b]
	\centering
		\includegraphics[width=\columnwidth]{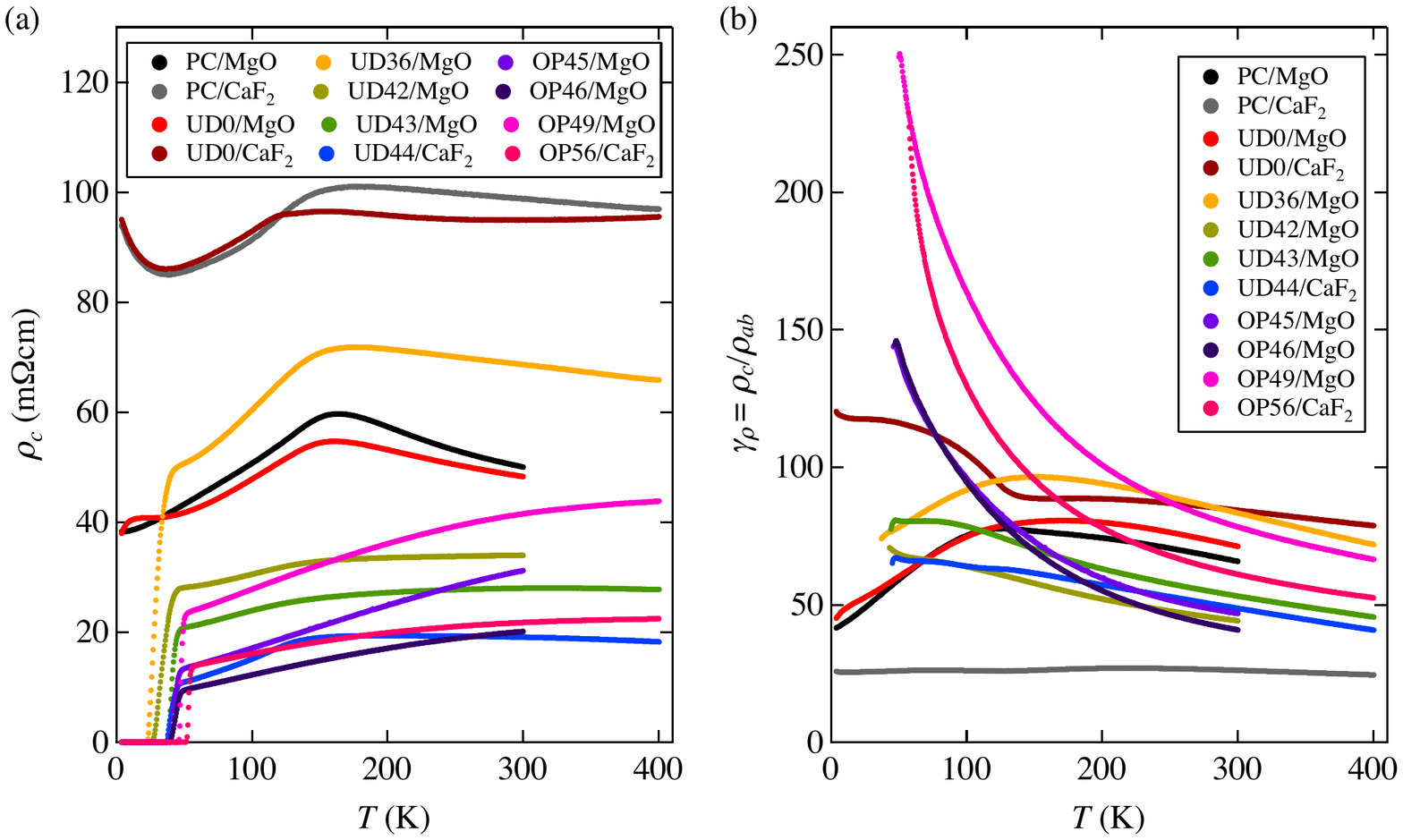}
		\caption{(a) The $c$-axis resistivity curves for all samples tabulated in table\,\ref{tab:table1}. (b) The temperature-dependency of resistivity anisotropy
		 $\gamma_{\rho}(T)=\rho_{c}/\rho_{ab}$ for the films presented in (a).} 
\label{fig:figure4}
\end{figure}

The $c$-axis resistivity $\rho_{c}$ can be calculated by using eq. (1) and the growth angle $\alpha$ [fig.\,\ref{fig:figure4}(a)]. For the parent NdFeAsO and under-doped NdFeAs(O,F) films, a kink in the temperature dependence of $\rho_{c}$ due to the structural transition was observed at around 150\,K. Below the structural transition temperature ($T_{\rm str}$), $\rho_{c}$ decreased with lowering $T$. Here, a distinct feature was observed for the films on CaF$_2$ substrates: For both PC/CaF$_2$ and UD0/CaF$_2$ $\rho_{c}$ started to increase at around 30\,K with decreasing temperature. Such behaviour was not observed for the corresponding films grown on MgO substrates. For all superconducting UD samples $\rho_{c}(T)$ decreased below $T_{\rm str}$, and a gradual decrease of $d\rho_{c}/dT$ close to $T_{\rm c}$ was recognised. Similar behaviour was observed in other FBS single crystals like Ba$_{0.81}$K$_{0.19}$Fe$_2$As$_2$, BaFe$_2$(As$_{0.77}$P$_{0.23}$)$_2$ (i.e. under-doped regime)\,\cite{Tantar-2}, and LiFeAs\,\cite{Song}. For all OP samples, on the other hand, the temperature-dependent $\rho_{c}(T)$ was metallic, which differs from the reports on PrFeAsO$_{0.7}$\,\cite{Kashiwaya}, SmFeAs(O,F)\,\cite{Moll} and SmFeAsO$_{0.9}$H$_{0.1}$\,\cite{Iimura} single crystals. To identify the reason of this difference, it is desired to measure $\rho_{c}(T)$ using NdFeAs(O,F) single crystals. However, the size of available single crystals is limited. Further investigation is necessary.

In fig.\,\ref{fig:figure4}(b) the temperature-dependent resistivity anisotropy $\gamma_{\rho}(T)=\rho_{c}/\rho_{ab}$ for all samples is shown. 
Again distinct features were observed for PC/CaF$_2$ and UD0/CaF$_2$: 1) $\gamma_{\rho}$ for PC/CaF$_2$ was almost constant irrespective of $T$. On the other hand for PC/MgO and UD0/MgO $\gamma_{\rho}$ increased with decreasing $T$ down to $T_{\rm str}$ and decreased thereafter. A similar beahviour was also observed for UD36/MgO. 2) For UD0/CaF$_2$ $\gamma_{\rho}$ increased with decreasing $T$ down to $T_{\rm str}$, which is similar to UD0/MgO, but increased further rather rapidly below $T_{\rm str}$, which is different from UD0/MgO.

For the under-doped superconducting samples except UD36/MgO, $\gamma_{\rho}$ increased monotonously with decreasing $T$ down to $T_{\rm c}$ irrespective of the substrate. The increase was especially strong for the OP series samples, and showed an exponential-like $T$ dependence. For OP49/MgO and OP56/CaF$_2$, $\gamma_{\rho}$ were more than 220 near $T_{\rm c}$, which were quite large values compared to single crystals. Note that $\rho_{c}(T)$ was metallic for both OP49/MgO and OP56/CaF$_2$, and the increase of $\gamma_{\rho}(T)$ with lowering $T$ is because the rate of decrease of $\rho_{ab}(T)$ was much faster than $\rho_{c}(T)$.

\begin{figure}[ht]
	\centering
		\includegraphics[width=\columnwidth]{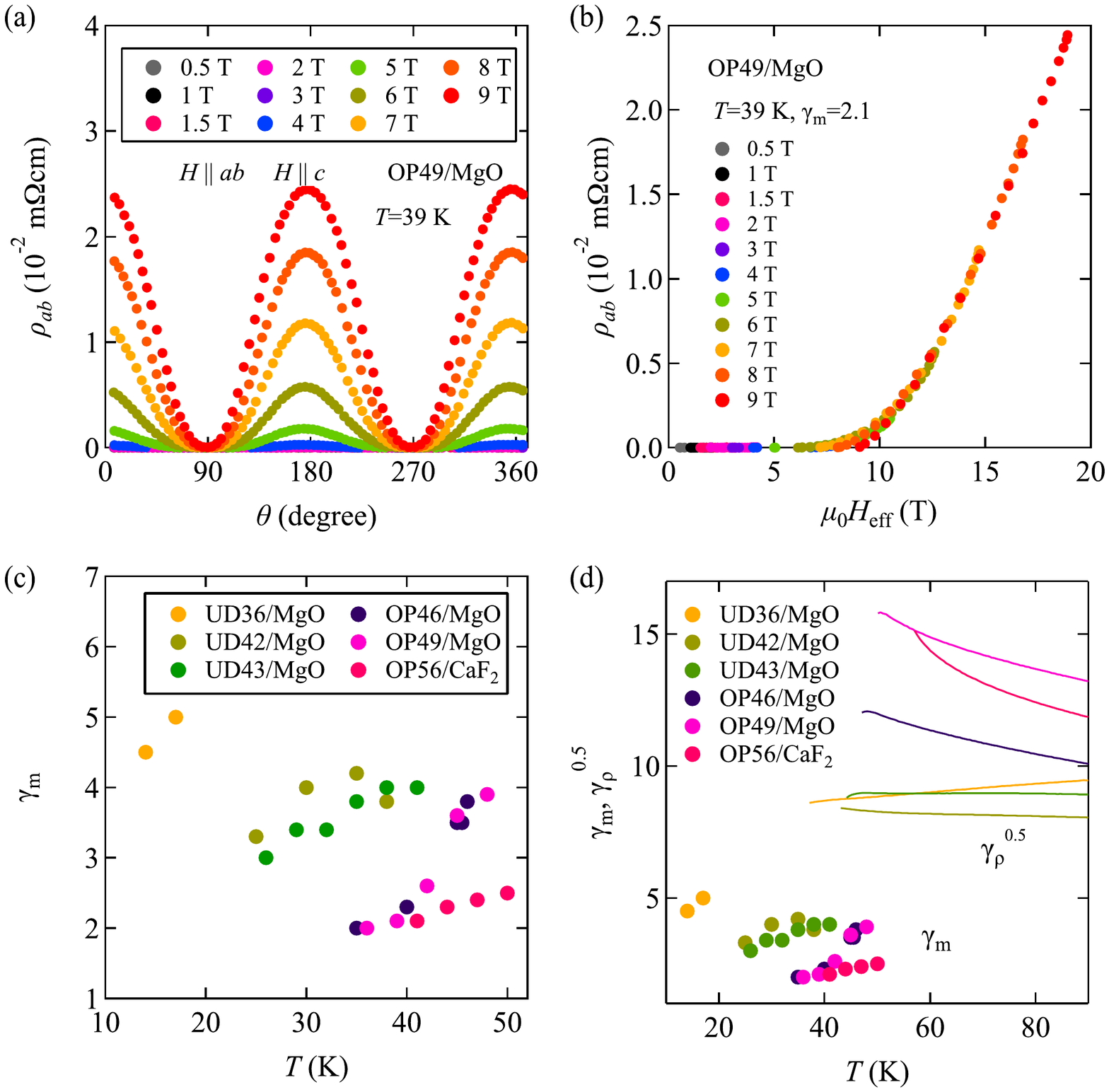}
		\caption{(a) Angular dependence of $\rho_{ab}$ for the NdFeAs(O,F) film OP49/MgO at 39\,K in various magnetic fields. (b) The scaling behaviour of $\rho_{ab}(\theta)$ as a function of $H_{\rm eff}$. (c) The mass anisotropy $\gamma_m$ derived from the anisotropic Ginzburg-Landau approach as a function of temperature for all superconducting films. (d) The plots of the square root of $\gamma_{\rho}$($T$) shown in fig.\,\ref{fig:figure4}(b) and $\gamma_m(T)$.} 
\label{fig:figure5}
\end{figure}

In order to evaluate the mass anisotropy in the superconducting state, the angular dependence of resistivity $\rho_{ab}(\theta)$ was measured 
below $T_{\rm c}$. Figure\,\ref{fig:figure5}(a) shows the results for OP49/MgO measured at 39\,K in various magnetic fields. On the assumption that no correlated defects are present, the mass anisotropy governs the resistivity anisotropy. In this case, $\rho_{ab}(\theta)$ can be scaled with $H_{\rm eff}$ $[H_{\rm eff}=H\epsilon(\theta)$, $\epsilon(\theta)=\sqrt{\cos ^2\theta+\gamma_m^{-2}\sin ^2\theta}$], where $\gamma_m$ is the mass anisotropy\,\cite{Blatter}. This approach was applied to a NdFeAsO$_{0.82}$F$_{0.18}$ single crystal ($T_{\rm c}$=51.5\,K) and its $\rho_{ab}(\theta)$ at given temperatures were scaled with an appropriate $\gamma_m$, which increased with lowering $T$\,\cite{Jia-1}.
Near $T_{\rm c}$, $\gamma_m$ was 5.2, which was consistent with the value evaluated from the ratio of the upper critical field $H_{\rm c2}^{ab}/H_{\rm c2}^{c}$\,\cite{Jia-2}.

The scaling behaviour of $\rho_{ab}(\theta)$ for OP49/MgO  as a function of $H_{\rm eff}$ is shown in fig.\,\ref{fig:figure5}(b). It is clear that all data collapsed onto a single curve with $\gamma_ m$=2.1. We also evaluated $\gamma_m$ at different temperatures as well as for all superconducting films. As can be seen, the resulting $\gamma_m$ was found to decrease with decreasing temperature [fig.\,\ref{fig:figure5}(c)]. This observation differs from the result obtained from the single crystal\,\cite{Jia-1}. It is interesting to note that $\gamma_ m$ for the under-doped films are higher than those for the optimal-doped films. This may be due to the decrease in the interlayer coupling, which was similarly observed in cuprates\,\cite{Kishio}.

By assuming $\rho_{ab}=m^*_{ab}/ne^2\tau$ and $\rho_{c}=m^*_{c}/ne^2\tau$, where $n$ is the carrier density, $e$ the electric charge, 
and $\tau$ the relaxation time of carriers, the mass anisotropy $\gamma_m^2$ should be equal to $\gamma_ \rho$ at $T_{\rm c}$. The mass anisotropy and square root of $\gamma_{\rho}$ as a function of temperature are shown in fig.\,\ref{fig:figure5}(d). As can be seen, the temperature dependence of the anisotropy showed a discontinuous change at $T_{\rm c}$. The difference between $\gamma_\rho^{0.5}$ and $\gamma_m$ is getting larger with increasing the F content. Similar discrepancy of the anisotropy at $T_{\rm c}$ was observed in YBCO. The resistivity anisotropy obtained from a vicinal YBCO film at 100\,K was around 60\,\cite{Emergo}, which is in good agreement with the value obtained from twin free single crystals\,\cite{Friedmann}. On the other hand, $\gamma_m$ obtained from the anisotropic Ginzburg-Landau approach for clean YBCO films was around 5\,\cite{Civale}, which differs from the values derived from the resistivity anisotropy. Albeit the reason why the mass anisotropy is different in the normal and superconducting states for YBCO is unclear, such difference may be possible for FBS due to their multi-band nature. If the dominant electronic bands which govern the transport properties are different in the normal and superconducting states, the mass anisotropy could be different in the two states. This may partially explain our experimental results.

\section{Summary} 
NdFeAs(O,F) epitaxial thin films having different F contents were grown successfully on vicinal cut MgO and CaF$_2$ substrates by MBE. 
Using these films the anisotropy of the electrical transport properties were investigated. The $c$-axis resistivity was always higher than $\rho_{ab}$ irrespective of the F content, resulting from the anisotropic electronic structure. The temperature-dependent $\rho_{c}(T)$ for superconducting films was metallic at low temperature, which differs from the other reports on the $Ln$FeAsO system. In the superconducting state, the mass anisotropy for the superconducting thin films were derived using the anisotropic Ginzburg-Landau approach. Near $T_{\rm c}$, the resultant values are different to these obtained from the resistivity anisotropy, which may be due to the multi-band nature of NdFeAs(O,F).

\begin{acknowledgments}
This work was supported by the JSPS Grant-in-Aid for Scientific Research (B) Grant Number 16H04646 as well as JST CREST Grant Number JPMJCR18J4. 
\end{acknowledgments}

\newpage
\section{Supplementary Information}
\begin{figure}[b]
	\centering
		\includegraphics[width=\columnwidth]{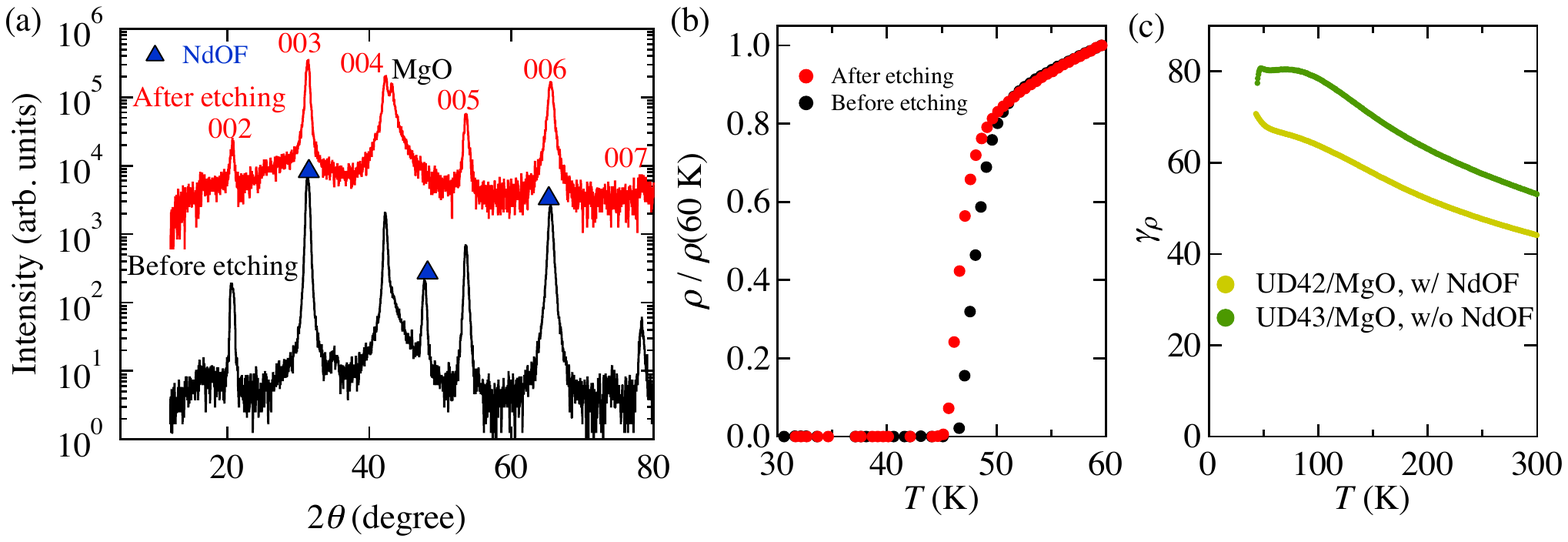}
		\caption{(a) The $\theta/2\theta$-scans and (b) the normalised resistivity curves of NdFeAs(O,F) grown on a 5$^\circ$ vicinal cut MgO substrate (OP49/MgO). The indices in (a) are for the NdFeAs(O,F) phase. (c) The temperature dependence of $\gamma_{\rho}$ for UD42/MgO and UD43/MgO.} 
\label{fig:figureS1}
\end{figure}

\beginsupplement
To investigate whether the presence of NdOF affects the superconducting properties, the NdOF over-layer was removed by Ar ion-beam etching. It is clear from fig.\,\ref{fig:figureS1}(a) that the diffraction intensity from NdOF is significantly reduced after the etching process. After the etching process, the superconducting transition temperature slightly reduced, but the difference was less than 1\,K [fig.\,\ref{fig:figureS1}(b)]. Figure\,\ref{fig:figureS1}(c) shows the temperature dependence of $\gamma_{\rho}$ for UD42/MgO and UD43/MgO. Here, the NdOF over-layer was removed for UD43/MgO, whereas NdOF was present for UD42/MgO. As can be seen, the temperature-dependent $\gamma_{\rho}$ for both films showed almost the same behaviour. From these results, it is clear that the superconducting properties are not affected significantly by the presence/absence of the NdOF over-layer.

\begin{figure}[t]
	\centering
		\includegraphics[width=\columnwidth]{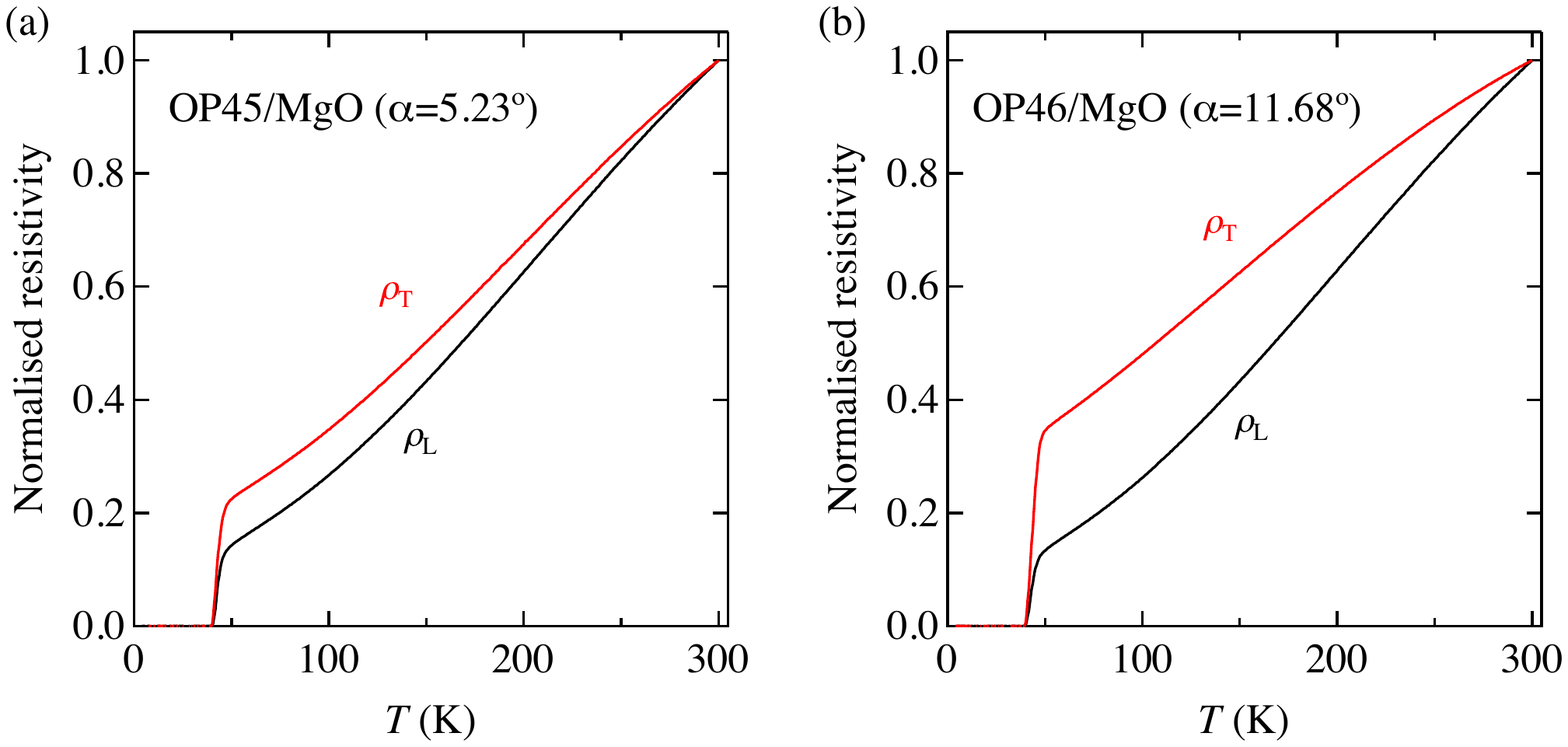}
		\caption{The normalised resistivity curves for (a) OP45/MgO and (b) OP46/MgO.} 
\label{fig:figureS2}
\end{figure}

 Figure\,\ref{fig:figureS2} compares the normalised resistivity for OP45/MgO and OP46/MgO. As can be seen, the resistivity difference between $\rho_{\rm T}$ and $\rho_{\rm L}$ for OP46/MgO is larger than that for OP45/MgO due to the larger vicinal angle, whilst $\rho_{\rm L}$ for both samples are almost identical to each other. The respective $\rho_{\rm L}$(=$\rho_{ab}$) values at 300 K for OP45/MgO and OP46/MgO were 0.67 m$\Omega$cm and 0.5 m$\Omega$cm. Hence, the superconducting properties are not affected by the vicinal angle.

\end{document}